\documentclass{emulateapj}

\usepackage[normalem]{ulem}
\usepackage{color}

\newcommand{\htwo}{\mbox{\rm H$_2$}}

\newcommand{\halpha}{\mbox{\rm H$\alpha$}}

\slugcomment{Accepted for publication in ApJ}

\shorttitle{Dense Gas in the Antennae Galaxies}
\shortauthors{Bigiel et al.}

\begin{document}

\title{Dense Gas Fraction and Star Formation Efficiency Variations in the Antennae Galaxies}

\author{F.~Bigiel\altaffilmark{1}, A.K.~Leroy\altaffilmark{2,3}, L.~Blitz\altaffilmark{4}, A.D.~Bolatto\altaffilmark{5}, E.~da~Cunha\altaffilmark{6,7}, E.~Rosolowsky\altaffilmark{8}, K.~Sandstrom\altaffilmark{9,10}, A.~Usero\altaffilmark{11}}

\altaffiltext{1}{Institut f\"ur theoretische Astrophysik, Zentrum f\"ur Astronomie der Universit\"at Heidelberg, Albert-Ueberle Str. 2, 69120 Heidelberg, Germany; bigiel@uni-heidelberg.de}
\altaffiltext{2}{Department of Astronomy, The Ohio State University, 140 W 18$^{\rm th}$ St, Columbus, OH 43210, USA}
\altaffiltext{3}{National Radio Astronomy Observatory, 520 Edgemont Road, Charlottesville, VA 22903, USA}
\altaffiltext{4}{Department of Astronomy, Radio Astronomy Laboratory,
University of California, Berkeley, CA 94720, USA}
\altaffiltext{5}{Department of Astronomy and Laboratory for Millimeter-Wave Astronomy, University of Maryland, College Park, MD 20742, USA}
\altaffiltext{6}{Max-Planck-Institut f{\"u}r Astronomie, K{\"o}nigstuhl 17,
69117 Heidelberg, Germany}
\altaffiltext{7}{Centre for Astrophysics and Supercomputing, Swinburne University of Technology, Hawthorn, Victoria 3122, Australia}
\altaffiltext{8}{Department of Physics, University of Alberta, Edmonton, AB, Canada}
\altaffiltext{9}{Center for Astrophysics and Space Sciences, Department of Physics, University of California, San Diego, 9500 Gilman Drive, La Jolla, CA 92093, USA}
\altaffiltext{10}{Steward Observatory, University of Arizona, 933 North Cherry Avenue, Tucson, AZ 85721, USA}
\altaffiltext{11}{Observatorio Astronomico Nacional, Alfonso XII 3, 28014, Madrid, Spain}


\begin{abstract}
We use the CARMA millimeter interferometer to map the Antennae Galaxies (NGC\,4038/39), tracing the  bulk of the molecular gas via the $^{12}$CO(1-0) line and denser molecular gas via the high density transitions HCN(1-0), HCO$^{+}$(1-0), CS(2-1), and HNC(1-0). We detect bright emission from all tracers in both the two nuclei and three locales in the overlap region between the two nuclei. These three overlap region peaks correspond to previously identified ``supergiant molecular clouds.'' We combine the CARMA data with {\em Herschel} infrared (IR) data to compare observational indicators of the star formation efficiency (SFR/H$_2 \propto\ $IR/CO), dense gas fraction (HCN/CO), and dense gas star formation efficiency (IR/HCN). Regions within the Antennae show ratios consistent with those seen for entire galaxies, but these ratios vary by up to a factor of 6 within the galaxy. The five detected regions vary strongly in both their integrated intensities and these ratios. The northern nucleus is the brightest region in  mm-wave line emission, while the overlap region is the brightest part of the system in the IR. We combine the CARMA and {\em Herschel} data with ALMA CO data to report line ratio patterns for each bright point. CO shows a declining spectral line energy distribution, consistent with previous studies. HCO$^+$ (1-0) emission is stronger than HCN (1-0) emission, perhaps indicating either more gas at moderate densities or higher optical depth than is commonly seen in more advanced mergers.
\end{abstract}



\section{Introduction}
\label{intro}

The influence of local physical conditions on how efficiently gas forms stars plays a key role shaping galaxies morphologically, kinematically, and  chemically. This has prompted many studies of both the Milky Way and other galaxies aimed at linking the gas abundance and physical properties of the gaseous ISM to the observed star formation rate. Such studies provide key information to understand the physical processes governing star  formation on small and large scales and serve as input and the essential point of comparison for simulations on galactic or cosmological scales.

Many of these studies have focused on the relationship between \htwo\ mass, traced by CO emission, and the recent star formation rate (SFR),  often traced using, e.g., \halpha, UV or infrared (IR) emission \citep[e.g.,][]{kennicutt98,bigiel08,bigiel11,leroy08,leroy13}.  Those studies find a basic scaling between the molecular gas traced by CO (typical densities $\sim 10^3$\,cm$^{\rm -3}$ and temperatures $\sim10-20$\,K) and the SFR. However, in detail there are important variations of the ratio between SFR and CO including more efficient star formation in compact, dense systems \citep[e.g.,][]{daddi10,genzel10,garcia12}, more efficient star formation in galaxy centers \citep[e.g.,][]{leroy13}, a dependence on local dynamics \citep{meidt13}, and large samples revealing global trends with stellar mass and morphology \citep[e.g.,][]{saintonge11}. The contrast between vigorous starbursts and main sequence galaxies has drawn particular attention \citep{gao04a,gao04b,kennicutt98}, but all of these studies show that physical conditions within the molecular gas change systematically across the universe and that these changes in turn affect the star formation process.

On scales of individual clouds in the Milky Way, stars actually form in a small part of a cloud with the highest surface and volume densities \citep[e.g.,][]{heiderman10,lada10}. At the distances of other galaxies, this high surface density (A$_{\rm V} \sim 10$, $\Sigma \ga 100$~M$_\odot$~pc$^{-2}$) gas is best picked out by studying transitions like HCN(1-0) or HCO$^+$(1-0). Due to  their higher dipole moments compared to CO, these lines have higher critical densities, around $10^5-10^6$\,cm$^{-3}$ compared to $\sim 10^3$~cm$^{-3}$ for CO. Accounting for radiative trapping lowers the critical density for exciting emission, so that these lines are tracers of gas with effective density $\sim 10^5$~cm$^{-3}$. Analogously, observations of higher-J CO lines have been used to trace gas at densities intermediate between that traced by low-J CO emission and high critical density tracers like HCN \citep[][and references therein]{yao03,mao10,ueda12,wilson12}. Focusing on studies comparing emission from high critical density tracers such as HCN to signatures of embedded star formation, these have found a roughly linear correlation that spans from Milky Way cloud cores \citep{wu05,wu10} to the integrated emission from star forming galaxies and (U)LIRGs \citep{solomon92,gao04b}, a range of ten orders of magnitude in luminosity. This correlation has been interpreted as entire galaxies and individual dense cloud cores having a fixed SFR per unit dense  gas mass, at least on average. In turn, this suggests that a main variation driving the changes in the overall star formation efficiency of H$_2$ may be largely the fraction of the gas mass in this dense phase.

\begin{figure*}
\epsscale{1.1}
\plotone{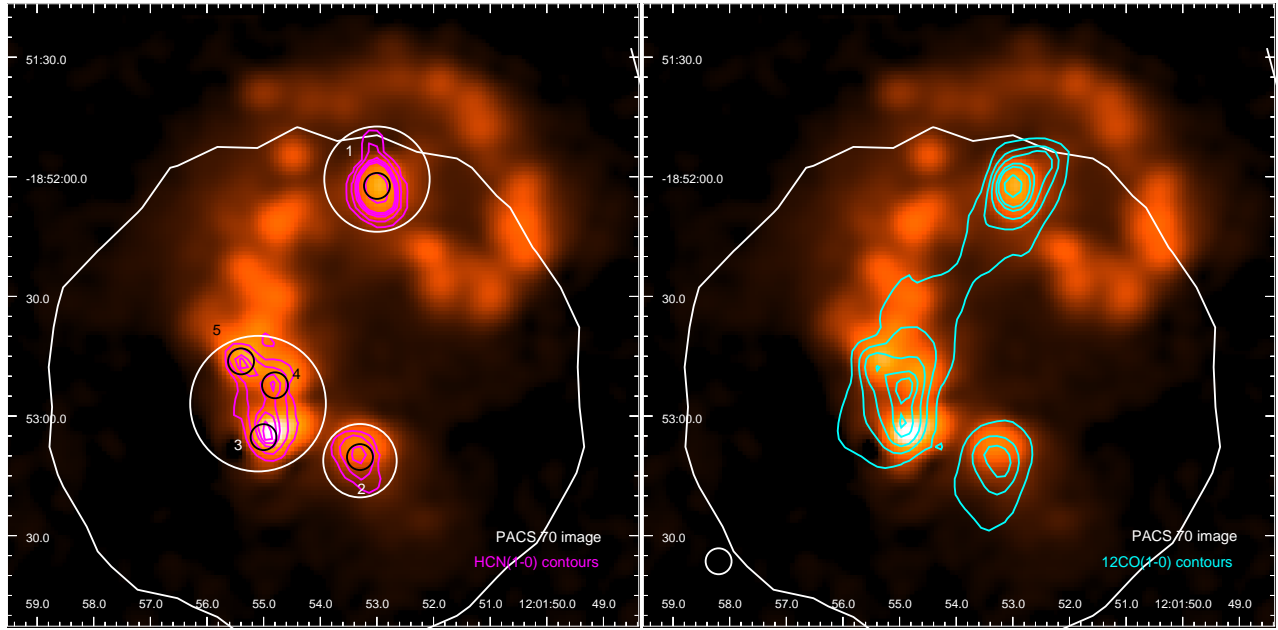}
\caption{CARMA maps of HCN ({\em left}) and CO ({\em right}) emission plotted in contour over the PACS 70$\mu$m image from the {\em Herschel} SHINING Survey (shown on a square-root stretch from 0-6000 MJy\,sr$^{-1}$). The large white circle shows the approximate good-sensitivity coverage for both molecular line maps. All data are at our common working resolution of 6.5\arcsec\ (compare beam in the right panel). In both panels the contours show the integrated intensity after masking the line data cube (see Section \ref{data}) and integrating to create an integrated intensity map. In the left  panel, HCN contours show 0.4, 2, 5, 6.5, 8, 17 K~km~s$^{-1}$. In the CO panel, the contours show 20, 100, 170, 255, 300, 420 K~km~s$^{-1}$. The five peaks visible in HCN correspond to the brightest IR concentrations. These are labeled in the left panel (compare ``ID'' in Tables 1, 2). Shown are also the apertures used to determine luminosities for further analysis (beam-sized on HCN peaks: black, larger regions: white, compare Table 2 and text).}
\label{fig:carmair}
\end{figure*}

These observations prompt two questions: how does the fraction of gas that is dense depend on the host galaxy and local conditions within an individual galaxy? And is the rate at which stars form from dense gas really constant or does it vary systematically? The former quantity, the fraction of dense gas, is related to the observable intensity ratio $I_{\rm HCN}/I_{\rm CO}$. The latter quantity, the star formation efficiency of dense gas, is traced by the $I_{\rm IR}/I_{\rm HCN}$ ratio. Work by \citet{gao04b} and \citet{garcia12} indicate that the dense gas fraction is higher in starburst galaxies. Despite initial indications that the IR-to-HCN ratio was quite constant \citep[see also][]{zhang14}, other recent work, including studies by \citet{garcia12} and \citet{kepley14} indicate that this quantity, too, may show important environmental variations (see also Usero et al., accepted).

Much insight can be gained in these topics from resolved observations of galaxies, especially those that host extreme conditions not readily accessible in the Milky Way \citep[e.g.,][]{paglione97}. With this in mind, this paper presents new Combined Array for Research in Millimeter-wave Astronomy (CARMA) observations of the high effective density transitions HCN(1-0), HCO$^+$(1-0), HNC(1-0), and CS(2-1) and of $^{12}$CO(1-0) emission in the Antennae galaxies (NGC\,4038/4039). We use these tracers to study dense gas fractions and scaling relations, placing individual regions within this nearest major galaxy merger onto the scaling relations discussed above. The relationships among the various lines also suggest variations in physical conditions from place to place within the Antennae.

The Antennae are the nearest major galaxy merger, at a distance of $\approx 22$~Mpc \citep{schweizer08}. The colliding galaxies exhibit bright CO emission \citep{stanford90,wilson00,wilson03,wei12,ueda12,whitmore14} at each of the two nuclei and in the ``overlap'' region where the two systems are colliding. The collision has led to abundant star formation and numerous young stellar clusters \citep[e.g.,][]{whitmore99}. The proximity of the galaxies and the extreme environment induced by the galaxy merger make the Antennae an ideal target for spatially resolved dense gas observations. \citet{gao01} made a first study of HCN in the Antennae using single-dish observations. These exclude much of the overlap region where most of the star formation takes place, however, and their resolution of 72\arcsec\ is more than a factor of 10 coarser than that of our new CARMA observations. With improved resolution and a field of view covering the entire system, we are able to study  individual regions within the galaxies.

\begin{deluxetable*}{ccccccccccll}
\tablewidth{0pc} 
  \tablecaption{Integrated intensities at matched angular resolution for each transition for the five HCN peaks labeled in Figure \ref{fig:carmair}}
  \tablehead{
  \colhead{Region} & \colhead{R.A.} & \colhead{Dec.} & \multicolumn{7}{c}{Integrated Intensity at HCN Peak} &  \colhead{IR Intensity} & \colhead{Notes} \\
    \colhead{ID}   & \multicolumn{2}{c}{(J2000)} & \colhead{$^{12}$CO} & \colhead{$^{12}$CO} & \colhead{$^{12}$CO} & \colhead{HCN} & \colhead{HCO$^{+}$} & \colhead{CS} & \colhead{HNC} & \colhead{70$\,\micron$} & \colhead{} \\ 
 \colhead{}   & \colhead{} & \colhead{} &  \colhead{(1-0)} & \colhead{(2-1)} & \colhead{(3-2)} & \colhead{(1-0)} & \colhead{(1-0)} & \colhead{(2-1)} & \colhead{(1-0)} &\colhead{} &\colhead{}\\     
\colhead{}  & \colhead{} & \colhead{} & \multicolumn{7}{c}{[K\,km\,s$^{-1}$]} & \colhead{[MJy\,sr$^{-1}$]} & \colhead{}}
 \startdata
1 &12:01:53.03  & -18:52:02.3  & 502.0 & 223.4 & 146.6 & 27.7 & 28.4 & 11.5  & 14.4 &  2799.9 & Northern Nucleus\\
2 &12:01:53.32  & -18:53:10.3  & 207.6 & 109.9 & 46.6  & 7.8  & 10.8 & 5.3   & 3.2  &  1049.7  & Southern Nucleus  \\
3 &12:01:54.93  &-18:53:05.3   & 265.8 & 142.6 & 72.8  & 8.7  & 14.5 & 8.2   & 6.3  &  6034.1 & Overlap, SGMC 4-5 \\
4 &12:01:54.80  &-18:52:52.3   & 325.4 & 114.7 & 51.0  & 10.6 & 16.9 & 7.0   & 6.0  &  1779.7 & Overlap, SGMC 2 \\
5 &12:01:55.36  & -18:52:46.3  & 239.9 & 88.6  & 50.7  & 9.0  & 17.3 & 7.5   & 6.1  &  3046.4 & Overlap, SGMC 1 
\enddata
\tablecomments{All line intensities come from CARMA except for $^{12}$CO(2-1) and $^{12}$CO(3-2), which we measure from ALMA Science Verification data. The 70$\mu$m intensity comes from the SHINING Survey with {\em Herschel}. SGMC refers to the supergiant molecular complex designation by \citet{wilson00}. The velocity windows for the 5 regions within which the intensities are integrated are 1450-1750, 1500-1750, 1400-1700, 1350-1700 and 1300-1600\,km\,s$^{-1}$, respectively.}
\label{table1}
\end{deluxetable*}

\section{Data}
\label{data}

\subsection{Observations}

We used CARMA to observe HCN(1-0), HCO$^+$(1-0), HNC(1-0) and CS(2-1) emission from the Antennae galaxies during 2011. We observed a 7-point hexagonal mosaic that covered the two nuclei and the overlap region using CARMA's C and D configurations ($\sim$16\,hr and $\sim$26\,hr total observing time, respectively). The FWHM of the final synthesized beam is $\sim5.7\times4.0\arcsec$ (for ${\rm HCN(1-0)}$), corresponding to $\sim500$\,pc at our adopted distance of 22\,Mpc. We covered each line with a 250\,MHz ($\sim830$\,km\,s$^{-1}$) wide band that had native channel width of $\sim$1.3\,MHz ($\sim$4.3\,km\,s$^{-1}$), which we smoothed to  8\,km\,s$^{-1}$ for all subsequent analysis. The 1$\sigma$ sensitivity per channel in the final data cubes at our working resolution of 6.5\arcsec (see below) is typically 6\,mJy\,beam$^{-1}$ ($\sim$22\,mK).

For complex gain and phase calibration, we observed the quasar 1130-148 every $\sim20$\,min. 3C273 was observed for bandpass calibration. We set the flux scale either by setting the flux of the secondary calibrator based on recent measurements in the CARMA calibrator flux archive (CALFIND\footnote{http://carma.astro.umd.edu/cgi-bin/calfind.cgi}) or by using planet observations combined with a planetary model. The two approaches yield results consistent within 15\%, which we adopt as the uncertainty on the flux scale. The data reduction was carried out using MIRIAD.

To provide a view of the total molecular gas reservoir that complements the high critical density transitions, we observed $^{12}{\rm CO(1-0)}$ using CARMA in its C (2009) and D configurations (2011) for a total of  30\,hr and 7.5\,hr, respectively. The D configuration observations were carried out using the same setup and calibrators that we used for the high critical density transitions. The older C configuration observations used a correlator mode with a channel width of $\sim1$\,MHz ($\sim$2.6\,km\,s$^{-1}$) and a bandwidth of 62\,MHz ($\sim$160\,km\,s$^{-1}$), with two spectral windows combined to cover the line. The data reduction followed the same procedure described  above. The synthesized beam of the combined C+D configuration CO cube is $\sim3.1\times2.3\arcsec$ (FWHM), or $\sim$300\,pc at the distance of the Antennae galaxies. At its native resolution, the resulting cube is roughly a factor of two deeper and higher resolution than the OVRO $^{12}{\rm CO(1-0)}$ observations of \citet{wilson00,wilson03}, though the two maps do not cover exactly the same area. The final data cube (at $6.5\arcsec$ resolution) has a sensitivity (1$\sigma$) of $\sim$25\,mJy\,beam$^{-1}$ ($\sim$54\,mK) in each 5\,km\,s$^{-1}$ wide channel.

We compare our results to ALMA observations of higher-$J$ CO lines and {\em Herschel} mapping of the dust continuum. From ALMA, we use science verification observations\footnote{http://almascience.eso.org/alma-data/science-verification} of $^{12}{\rm CO(2-1)}$ and $^{12}{\rm CO(3-2)}$. We performed some additional self-calibration on these data that moderately improved the dynamic range of the images, but this reprocessing does not affect the results of this paper. As an indicator  of recent star formation, we use the publicly available {\em Herschel} (PACS) 70\micron\ intensity map obtained as part of the SHINING Survey (P.I.: E.~Sturm).

We convolve all data sets to a common circular Gaussian beam with a FWHM of 6.5\arcsec, $\sim$700\,pc at the distance of the Antennae. For the Herschel PACS map we use the kernel from \citet{aniano11} to convert from the PACS map to a Gaussian PSF. We put all cubes and maps on the astrometric grid of the CARMA HCN data set,  which has 1\arcsec pixels.

\subsection{Measurements}

We detect each transition towards both nuclei and the overlap region (see Figure \ref{fig:carmair} and Table \ref{table1}). We quantify the line emission in two ways. First, we measure integrated intensities for all lines along lines of sight towards the five bright peaks. Second, we measure the luminosities for individual bright regions, the whole overlap region, and our whole map from integrating data cubes masked to contain only significant emission.

In Table 1 we report integrated intensities for all lines towards the five lines of sight that show the brightest HCN emission. These lines of sight correspond to the two galaxy nuclei and three previously known supergiant molecular clouds in the overlap region. We measure the intensities from the $6.5\arcsec$ common-resolution data cubes by integrating the line emission, in Kelvins, across a velocity window selected by hand to encompass all bright line emission. Typically this line width is about 300\,km\,s$^{-1}$. For these measurements, no additional masking is applied beyond the choice of velocity window. For these integrated intensities, we calculate statistical uncertainties of $\approx 1$~K~km~s$^{-1}$ for the 90 GHz lines and $\approx 2$\,K\,km\,s$^{-1}$ for $^{12}{\rm CO(1-0)}$. We also estimate an uncertainty of $\approx 20\%$ due to choice of methodology; i.e., choice of integration window and a systematic gain calibration uncertainty of  $\sim15\%$. The Herschel data have a calibration uncertainty of $\sim5\%$ according to the PACS Observer's Manual. This is almost always larger than the uncertainties due to statistical noise.

We also measure $^{12}{\rm CO(1-0)}$ and HCN(1-0) luminosities from masked data cubes and report these values in Table 2. We derive these from integrated intensity maps produced by integrating signal-to-noise masked data cubes along the velocity axis. We produce one three-dimensional mask for the 90 GHz lines and one for CO. To do so, we select regions of the cube that have a signal-to-noise ratio greater than 3.5 (5) for the HCN or HCO$^{+}$ (CO) over at lease two adjacent velocity channels. We remove small regions (those less than a beam size in volume) because these likely represent statistical fluctuations. We then apply the mask to the cube and integrate to produce a map of integrated intensity. From these integrated intensity maps, we measure the flux summed in several different regions: beam-sized, 6.5\arcsec-diameter apertures centered on the HCN peak positions from Table \ref{table1}; larger circular apertures that encompass the entire northern nucleus, southern nucleus, and overlap region; and a sum over the whole map. Figure \ref{fig:carmair} shows the apertures and Table 2 reports the aperture positions and sizes and the luminosities derived from the measured fluxes and our adopted distance.

Our interferometer measurements with CARMA may resolve out emission on large spatial scales and the need to mask may cause us to miss low signal to noise emission. To check how much flux we may miss, we compare our luminosities (Table 2) to the  single dish measurements presented in \citet{gao01}. Accounting for differences in the adopted distance, we recover about 70\% and 50\% of the total $^{12}$CO(1-0) and HCN(1-0) luminosity, respectively, found by \citet{gao01}. Note that the \citet{gao01} CO observations cover the entire system and thus a significantly larger area compared to our map (see their Figure 2); from earlier work, we know that a significant amount of CO emission arises from the northern arm that is not included in our coverage. The flux in our CARMA CO map is about $25\%$ higher than the integrated result from the earlier OVRO (interferometer) work by \citet{wilson03}. Overall, missing flux does not appear to be an enormous concern for the CO.

For the HCN, the amount of flux missed is less clear: the discrepancy with the single dish result reflects some combination of calibration uncertainties in both telescopes, statistical noise in the \citet{gao01} measurement, the limited coverage of our map, and flux missed due to the need to clip the data based on signal-to-noise. There is not a strong reason to expect the 90~GHz lines to be more spatially filtered than CO; they were observed with the same two CARMA configurations and to first order the structure of the galaxy should be similar in these two tracers of the molecular gas. The \citet{gao01} luminosity is computed from two pointings, one of which would include the northern arm not covered in our observations. Those measurements also have comparatively low signal to noise, as do our own data.

For the most part the faint thermal radio continuum at 90~GHz remains undetected. However, the HCN and HCO$^+$ spectra in region ID 3 do show some hints of thermal continuum emission, though at low-significance. We remove this component during data processing.

\section{Results and Discussion}
\label{results}
\subsection{HCN-Selected Lines-of-Sight}
\label{los}

\begin{figure*}
\epsscale{1.1}
\plotone{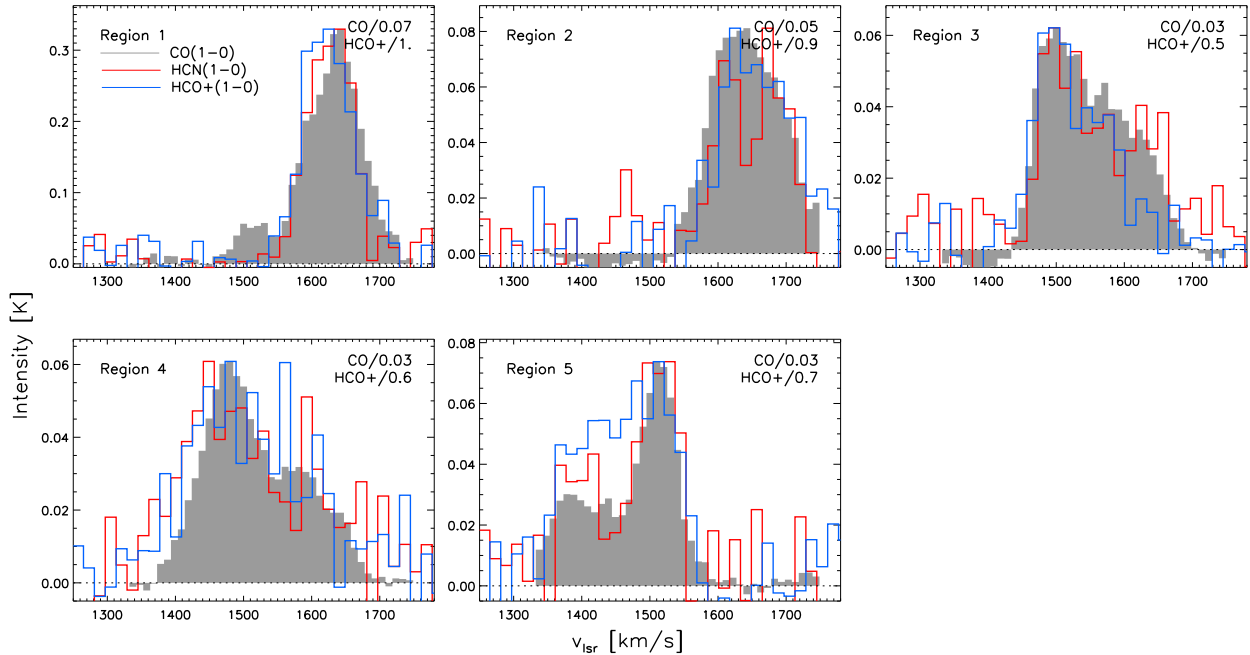}
\caption{$^{12}$CO(1-0) (gray), HCN(1-0) (red) and HCO$^{+}$ (blue) spectra for the 5 HCN-selected lines of sight (see main text). The spectra are derived from our common 6.5\arcsec\,resolution, unmasked data cubes and plotted at 16 (HCN, HCO$^{+}$) and 10~km~s$^{-1}$ (CO) resolution, respectively. Regions 1 and 2 are the northern and  southern nucleus. The other lines-of-sight are in the overlap region. The CO and HCO$^{+}$ spectra are scaled to match the HCN peak with the scaling factor given in the panels. We detect HCN at high signal-to-noise for each line-of-sight and the spectra and linewidths closely resemble each other.}
\label{fig:spec}
\end{figure*}

Our observations cover both nuclei and the overlap region and following \citet{wilson00,wilson03} we detect abundant molecular gas across the galaxies, with concentrations at both nuclei and a large mass of molecular gas in the overlap region. Figure \ref{fig:carmair} shows our new $^{12}$CO(1-0) map in contour. The emission from the overlap region is brightest in three peaks labeled ``supergiant molecular complexes'' (SGMCs) by \citet{wilson00}. Figure \ref{fig:carmair} also shows HCN intensity contours. The faintness of the HCN line compared to CO means that we detect HCN at high signal-to-noise ratio over a smaller area than CO with bright HCN mainly seen in the three SGMCs in the overlap region and the two nuclei. A similar picture emerges for the other high critical density transitions, with high signal-to-noise emission concentrated in these five regions: the two nuclei and three bright complexes in the overlap region \citep[SGMC \#1, \#2, and \#4+5 from][]{wilson00}. In Table \ref{table1}, we report integrated intensities of each of our high critical density lines and IR emission at 70$\mu$m for each of these five peaks. 



\begin{deluxetable*}{lcccccc}
\tablecaption{$^{12}$CO(1-0), HCN(1-0) and TIR luminosities for the colored data points in Figure \ref{fig:scaling}}
\tablehead{
  \colhead{Region} & \multicolumn{2}{c}{Position} & \colhead{Aperture} & \multicolumn{3}{c}{Luminosities}\\
   \colhead{}    &   \colhead{R.A.} & \colhead{Dec.} & \colhead{Diameter} & \colhead{12CO(1-0)} & \colhead{HCN(1-0)} & \colhead{TIR}  \\  
    \colhead{}     &  \multicolumn{2}{c}{J(2000)} & \colhead{[\arcsec]} & \multicolumn{2}{c}{[10$^7$\,K\,km\,s$^{-1}$\,pc$^{2}$]} & \colhead{[10$^9$\,L$_{\odot}$]}}     
 \startdata
  ID 1&   12:01:53.03   &  -18:52:02.3  & 6.5 &   16.4  &     0.81  &   2.5   \\
  ID 2&  12:01:53.32 &  -18:53:10.3  & 6.5 &    7.5    &   0.19  &   1.1    \\
  ID 3&  12:01:54.93  & -18:53:05.3  & 6.5 &   9.2    &    0.25  &   4.7    \\
  ID 4&  12:01:54.80  & -18:52:52.3 &  6.5 &   11.7    &    0.21  &   1.8    \\
  ID 5&   12:01:55.36  &  -18:52:46.3  & 6.5 &   8.0    &    0.18   &  2.5    \\
  Nuc. N&  12:01:53.03  &  -18:52:00.6  & 26.4 &   79.8  &       2.08   &  8.8    \\
  Nuc. S&  12:01:53.32  &  -18:53:11.2  & 18.4 &   31.8   &     0.38  &   4.5    \\
  Overlap&  12:01:55.10  &  -18:52:56.9  & 34.0 &   113.0  &       1.41  &   28    \\
  System& \multicolumn{3} {c}{\nodata}&     264.8  &      3.87 &    69   
\enddata
\tablecomments{The total infrared luminosity estimate from 70$\mu$m intensity follows \citet{galametz13}. The line luminosities are derived in manually selected apertures from our signal-to-noise masked integrated intensity maps (see Section \ref{data}).We use circular, beam-sized apertures for the 5 HCN peaks from Table 1, 3 larger regions covering the nuclei and the overlap region and integrate all emission in our maps to characterize the system as a whole. For this integral in the 70 $\mu$m map, we only consider emission in the region covered by our CARMA observations (compare Figure 1, left panel). Uncertainties are $\approx 35\%$ for integrated line intensities (Section 2) with an additional 10\% uncertainty for the distance measurement \citep{schweizer08} which should be taken into account (implying a 20\% uncertainty on any luminosity).}
\label{table2}
\end{deluxetable*}

From Figure \ref{fig:carmair} and Table \ref{table1} we see that the northern nucleus (ID 1, NGC\,4038) is the brightest region in all molecular lines, though not in infrared intensity. The southern nucleus (ID 2, NGC\,4039) is one of the two faintest regions in each tracer. Note that \citet{brandl09} do not detect signatures of AGN activity in either nucleus based on mid-IR spectroscopy so we do not expect AGN activity to account for a significant amount of the IR or dense gas emission appearing from the nuclei. Meanwhile, the brightest IR regions in the galaxy are not at either nucleus, but in the overlap region, where the emission is dominated by young stellar populations and active star formation \citep[e.g.,][determine the ages of massive clusters in this region to be a few Myr]{whitmore99} and kinematic signatures indicate possible collisions between massive clouds and infalling gas \citep[regions D and E in][]{wei12}. Our region \# 3, \citet{wilson00} SGMC \# 4-5, is the brightest IR source. In this region, the IR luminosity (and so the implied SFR) approaches the value for the entire Milky Way but compressed into a roughly kpc-sized region \citep{vigroux96,brandl09,klaas10}. A lack of detected X-ray emission and low radio continuum-to-IR ratios \citep[][and references therein]{klaas10}, detected water maser emission \citep{brogan10}, and low cluster ages \citep[e.g.,][]{zhang10} all  argue for this region \#3 being a young structure. 


Figure \ref{fig:spec} compares the line profiles for the bulk gas tracer $^{12}$CO(1-0) and the dense gas tracers HCN(1-0) and HCO$^{+}$(1-0). We scale the CO and HCO$^{+}$ by a factor chosen to match the peak of the HCN spectrum and bin the spectra to a velocity resolution of 16 (HCN, HCO$^{+}$) and 10~km~s$^{-1}$ (CO), respectively, to increase signal-to-noise. If one accounts for the substantially lower signal-to-noise of the HCN, the spectral profiles mostly agree between the three lines, suggesting that for the most part dense gas and the bulk of the molecular medium are well mixed in the different regions. There are a few notable exceptions, however; the low velocity weaker component seen in CO towards the northern nucleus (first panel) does not appear to have a counterpart in HCN or HCO$^{+}$, suggesting that perhaps this component is less dense than the gas giving rise to the main line. In the overlap region, our Region 5 also shows some variation between spectral components, with HCO$^+$ and HCN brighter relative to CO in the low velocity component. Throughout the overlap region, HCO$^{+}$ emission is observed to be stronger relative to HCN, which is discussed in Section \ref{ratios}, where we focus on molecular line ratio variations.

\subsection{Luminosity Ratios, Dense Gas Fractions, and the Star Formation Efficiency of Dense Gas}

\begin{figure*}
\epsscale{1.1}
\plotone{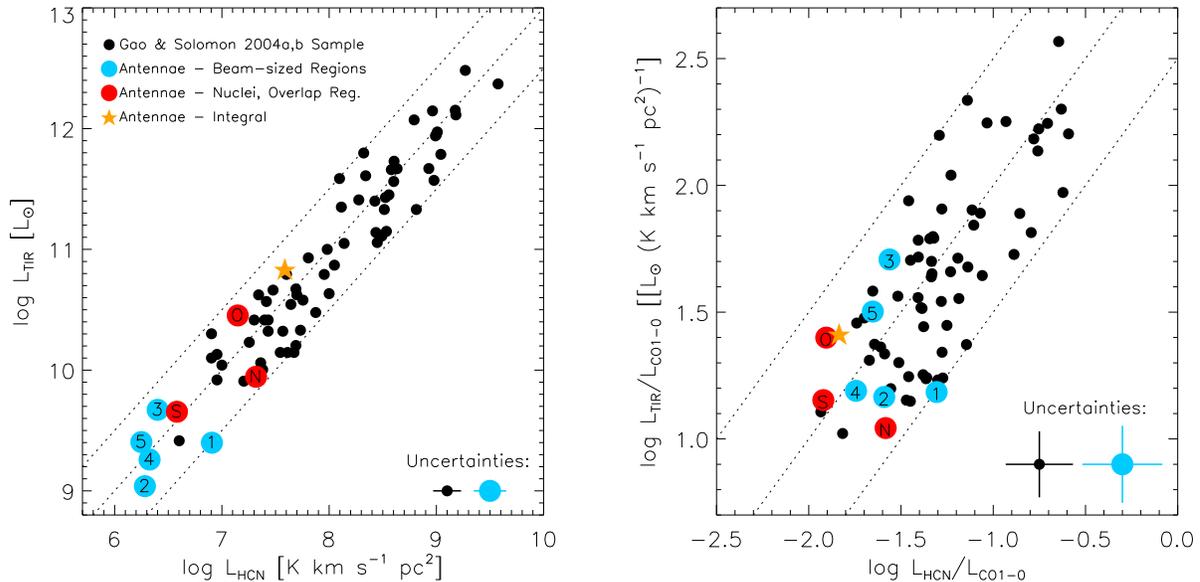}
\caption{({\em Left panel}) $L_{\rm TIR}$, IR luminosity, vs. $L_{\rm HCN}$, HCN luminosity. Black points show the (U)LIRG and disk galaxy sample from \citet{gao04b} (one point represents one galaxy).  Blue points show measurements for our five HCN-selected regions. Red points represent integrated luminosities for the entire overlap region and the two nuclei. The yellow star is the luminosity integral for the observed area of the Antennae (see Section 2 and Table \ref{table2}). The dotted lines show fixed IR-to-HCN ratios separated by a factor of three. Our resolved measurements extend the trend from \citet{gao04b}  to lower luminosities and we find that variations within the Antennae are of similar magnitude than the galaxy-to-galaxy scatter in the \citet{gao04b} sample. ({\em Right panel}) IR-to-CO vs. HCN-to-CO ratio as tracers of the star formation efficiency of the bulk molecular gas ($\sim$ IR-to-CO) and dense gas fraction ($\sim $ HCN-to-CO), respectively. The increase in the former can partly be explained by an increase in the latter, though there is substantial variation in the star formation efficiency of the dense gas ($\propto$ IR-to-HCN). Diagonal lines indicate constant ratios of IR-to-HCN separated by factors of three.}
\label{fig:scaling}
\end{figure*}

A main motivation for studying the high effective density lines are studies showing repeatedly a good overall correlation between IR luminosity and the luminosity of high effective density tracers. Emission from these tracers is presumed to trace the amount of dense gas \citep[following][]{solomon92,gao04b,wu05,wu10,juneau09,garcia12} while IR luminosity indicates the amount of recent star formation. In the left panel of Figure \ref{fig:scaling} we compare our measurements to the 65 (U)LIRGS and spiral galaxies from \citet{gao04b} (black points). We plot our measurements of the Antennae in three different (non-independent), ways as defined in Section 2 and reported in Table \ref{table2}. First, we show the luminosities for beam-sized regions centered on the HCN-selected lines-of-sight in blue (with regions labelled following Tables 1 and 2). In red, we show luminosities for the entire northern and southern nucleus as well as the entire overlap region (labeled ``N'', ``S'' and ``O'', respectively). Finally, in yellow we show the luminosity integrated over the whole part of the system that we observe.

The left panel of Figure \ref{fig:scaling} shows that our resolved measurements for individual regions extend the observed trend from  \citet{gao04b} towards lower luminosities. The dotted lines in the left panel of the Figure have a unity slope and indicate a factor of 3 scatter. The error bars represent an uncertainty of 35\% for our line luminosities, reflecting mostly gain (calibration) and methodology (see Section 2 and Table 2). We stress that in particular calibration uncertainties are strongly correlated, i.e., changes would affect all Antennae points in the same way. We adopt 30\% as the typical value for the \citet{gao04a} sample as quoted in their paper. We ignore the small uncertainty on the TIR luminosity as well as the correlated distance uncertainty; changing the distance would move all data points collectively along the dotted lines. Large regions of the Antennae galaxies populate this figure in regimes between entire galaxies and individual clouds. In this sense, our results resemble recent results mapping individual galaxies \citep[][Bigiel et al. in prep.]{kepley14}, integrating whole galaxies \citep{garcia12} and sampling parts of galaxy disks (Usero  et al., accepted). In each of these cases, the relationship between HCN (or HCO$^+$) and infrared luminosities at intermediate scales extends the relationship found for starbursts and bright whole galaxies (though we caution that the luminosity of a part of a galaxy is a somewhat arbitrary quantity).

Data like those in the left panel have motivated the idea that the mass fraction of dense gas drives the overall star formation efficiency of molecular gas, with star formation in dense gas as a fairly universal process \citep[e.g.,][]{gao04b,lada10}. A simple way to explore this hypothesis is to plot the observed ratio of IR-to-CO luminosity as a function of HCN-to-CO luminosity. The IR-to-CO ratio should, for dense systems, trace the efficiency with which the bulk molecular gas is converted to stars while the ratio of HCN-to-CO emission offers an observational tracer of the fraction of gas mass in a dense phase.

We re-plot the data in this space in the right panel of Figure \ref{fig:scaling}. \citet{gao04b} found that the HCN-IR correlation remains approximately linear  while the IR-to-CO ratio increases as one moves from fainter to brighter systems. The resulting correlation in IR-to-CO vs. HCN-to-CO space appears in the black points. However, just as striking as the existence of both correlations is the large scatter in both plots. Though in the right panel, HCN-to-CO coarsely predicts the IR-to-CO ratio, there is scatter of about a factor of 3 about a constant IR-to-HCN ratio (indicated by the dotted lines). If there were simply a fixed dense gas star formation efficiency with the  two quantities simply traced by IR and HCN emission, then the data would follow a diagonal line in this plot. This is clearly not the case. Despite the small number of independent regions in the Antennae, this already appears in our new data, which show a scatter comparable to the \citet{gao04b} data set. Within our small Antennae sample, knowing HCN-to-CO would not have helped to predict the IR-to-CO ratio, which traces the rate at which the bulk molecular gas forms stars.

We can also attempt to recast the observed IR, CO, and HCN  emission into more physical quantities with the caveat that such conversions come with substantial uncertainty. To  estimate approximate physical values for the total and dense molecular gas mass and star formation rate, we adopt  ${\rm \alpha_{HCN}=10\,M_{\odot}\,(K\,km\,s^{-1}\,pc^{2})^{-1}}$ \citep{gao04b}, a Galactic $\alpha_{\rm CO} = 4.35$~M$_{\odot}$~(K~km~s$^{-1}$~pc$^{2}$)$^{-1}$, and the TIR-to-star  formation rate conversion of \citet{murphy11}.  We note that the TIR-to-star formation rate conversion depends on the SF time scales and history and for very recent star formation the underlying assumption of continuous star formation over $\sim100$\,Myr time scales introduces substantial additional uncertainty. Using these conversions, however, the dense gas depletion time (${\rm \tau_{dense}=M_{dense}/{\rm SFR}}$) is $\sim$40\,Myr for the entire (observed) system, in excellent agreement with the value from  \citet{gao01}. For the northern, southern nucleus and overlap region separately, $\tau_{\rm dense} \sim$160, 60 and 36\,Myr, respectively. The depletion times for the CO-traced bulk molecular gas are $\sim$1.1\,Gyr for the entire system, and $\sim$2.6, $\sim$2.1 and $\sim$1.2\,Gyr, respectively, for the sub-regions.

The HCN-to-CO luminosity ratio shows substantial variation across the local galaxy population, ranging from a few percent to several tens of percent, \citep[e.g.,][Usero et al. accepted, and references therein]{sorai02,gao04b}. In the Antennae, we derive ${\rm L_{HCN}/L_{12CO(1-0)}}$ to be of the order of a few percent: $\sim1.5\%$ for the system as a whole, $\sim2.6\%$ for the northern nucleus, and around $1\%$ for the overlap region and the southern nucleus. Adopting the conversion factors above, the dense gas fractions by mass would be about twice these values, $\sim 3\%$ for the system as a whole. As pointed out by \citet{gao01}, this is a rather small dense gas fraction compared to what would be inferred for (U)LIRGs, which often show HCN-to-CO luminosity ratios $>\sim 0.06$ and dense gas mass fractions $>\sim 10\%$ \citep[see also][]{baan08}. Instead, this dense gas mass fraction is similar to that seen in Milky Way clouds \citep[e.g.][]{evans91,lada10}. This argues that, perhaps surprisingly, the high turbulent pressure in the Antennae has not dramatically changed the dense gas fraction in the molecular gas \citep[see discussion in][]{wilson03}. Here we extend this observation from a statement about the whole system to observations of individual regions within the Antennae galaxies.

The picture presented by these luminosity ratios reinforces that although the Antennae are rich in gas, the overlap region holding most of the gas may not yet be overwhelmingly dense. Meanwhile, the two nuclei show starkly different behavior. And perhaps surprisingly, although dense gas and star formation traced by IR emission are clearly correlated, knowing the luminosity of line emission from  dense gas does not perfectly indicate the recent star formation rate as traced by IR emission. Natural explanations for this lack of correspondence between IR/CO and HCN/CO would be strong variations in the ratio of HCN emission to dense, immediately star-forming gas mass (e.g., due to optical depth or temperature effects), evolutionary effects with current HCN emission indicating future star formation but only poorly indicating of where stars have already formed, or real variations in the ability of dense gas to form stars.

\subsection{Ratios Among Molecular Transitions}
\label{ratios}

\begin{figure}
\epsscale{1.2}
\plotone{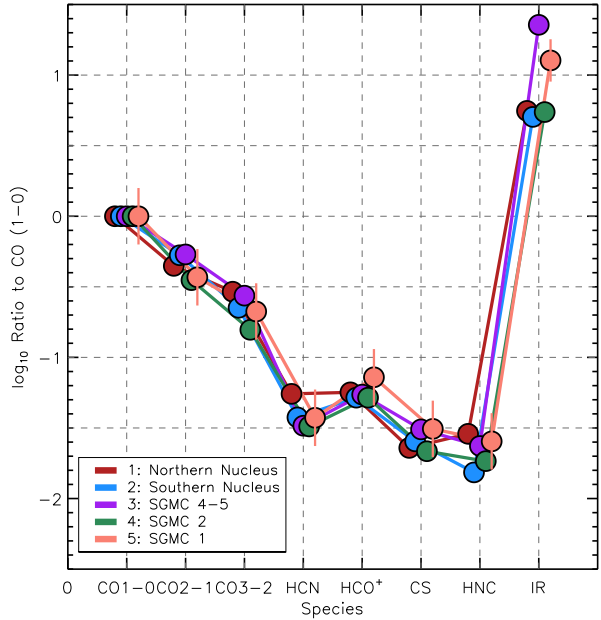}
\caption{Integrated line intensity ratios for all measured transitions and 70\micron\ emission from {\em Herschel} normalized to CARMA CO(1-0) for our five HCN-selected lines-of-sight (values from Table \ref{table1}). The CO(2-1) and CO(3-2) data are from ALMA, the other molecular line data from CARMA.}
\label{fig:linerats}
\end{figure}

We have so far considered $^{12}$CO(1-0) as a tracer of bulk gas and HCN(1-0) as a tracer of dense gas. We observe a suite of lines with high effective densities, HCO$^+$(1-0), CS(2-1), HNC(1-0). As part of its science verification program, ALMA has observed CO(3-2) and CO(2-1) emission from the Antennae galaxies. Table \ref{table1} and Figure \ref{fig:linerats} show the ratios among the integrated line intensities for these transitions, normalizing all measurements to CO(1-0). We plot one set of ratios for each region. The plotted errors (only shown for one set of measurements) assume a 35\% uncertainty on each individual integrated intensity measurement (see Section \ref{data} for details), though we again emphasize that these uncertainties are strongly correlated. We also caution that these ratios, in particular between ALMA and CARMA data, come with caveats due to different $u-v$ coverage and because they arise from interferometer-only data (see discussion in Section 2). Nonetheless some basic results are clear, with all measurements showing broadly similar line ratio patterns. CO emission (in K) declines from (1-0) to (2-1) to (3-2). The four high critical density tracers, HCN (1-0), HCO$^+$ (2-1), CS (2-1), and HNC (1-0) are all fainter, by more than an order of magnitude compared to CO (1-0) with the ordering usually HCO$^+$ (1-0) $>$ HCN (1-0) $>$ CS (1-0) $>$ HNC (1-0).

Our observed CO(3-2)/(1-0) ratio of $\sim 0.25$ and CO(2-1)/(1-0) ratio of $\sim 0.5$ could be reproduced (within $\approx 50\%$) either by cool, dense gas, in which the source function does not match the Rayleigh-Jeans approximation even at 230 and 345 GHz. Alternatively, they could be produced from hotter gas that has low enough density to sub-thermally excite the lines. With only the low-$J$ $^{12}$CO lines, these scenarios cannot be readily distinguished and a covariant set of densities and temperatures can reasonably reproduce our observed values in a one-zone model like RADEX \citep{vandertak07}. In fact, a combination of multiple physical conditions within the beam is almost certain \citep[see][for more discussion]{sliwa12,schirm14}. Indeed this would be required to simultaneously model both our high critical density lines and the CO emission; however, with only a single line per high-dipole species, this modeling introduces a free parameter (the abundance) for each new observation and does not yield significant insight.

As discussed above, the CO line ratios should be treated with some caution because the data come from different telescopes with different $u-v$ coverage. Indeed, the CO(3-2) / (1-0) ratio found here is much lower than that reported from single dish data by \citet{zhu03}. They found ratios near unity in the nuclei and varying with star formation activity level across the overlap region. Different spatial filtering or calibration differences between CARMA and ALMA might account for some of the discrepancy. The required sense is that the ALMA data recover a smaller fraction of the true CO flux than CARMA. Although ALMA is the more sensitive telescope and higher-$J$ CO emission might be more clumpy, the spatial filtering from ALMA is more severe. Some differential spatial filtering is certainly to be expected; the minimum projected baseline length for CARMA is $\sim2 k\lambda$ while for ALMA it is $\sim 12 k\lambda$ for CO (2-1) and $\sim 17 k\lambda$ for CO (3-2). These correspond to maximum recoverable spatial scales of $\sim100\arcsec$, $\sim18\arcsec$ and $\sim12\arcsec$ for CO (1-0), (2-1), and (3-2), respectively. However, spatial filtering occurs only in a single velocity channel and Figure \ref{fig:carmair} shows that the CO (1-0) and IR emission are already clumpy and concentrated towards the compact nuclei and the SGMCs, so we do not expect this effect to be severe. Indeed, after carefully matching $u-v$ coverage between SMA and OVRO data, \citet{ueda12} also find CO(3-2)/(1-0) ratios significantly less than reported by \citet{zhu03}. They find CO (3-2) / CO (1-0) ratios of 0.3--0.6 for essentially the same five regions considered here; we find $\sim 0.25$. Though clearly uncertain, our analysis supports the results of \citet{ueda12} and argues that the bulk of the gas in the Antennae is either cool or sub-thermally excited.

The ratios among the high effective density molecules are less systematically uncertain because they all come from one set of CARMA observations. These show that although we have discussed HCN (1-0) for ease of comparison to the literature, the HCO$^{+}$ (1-0) line is in fact brighter than HCN (1-0) across the Antennae. In this regard, the overlap region resembles M82 \citep{baan08,kepley14} but is weaker in HCN than is typical in the larger ($\sim 100$) sample of FIR-luminous galaxies studied by \citet{baan08}. Compared to the \citet{baan08} sample, the Antennae have a typical HCO$^{+}$/CO(1-0) ratio of 0.06 but the Antennae ratios of HCN/CO(1-0) $\sim 0.04$ and HNC/CO(1-0) $\sim 0.02$ are lower than average. The typical values of these ratios from \citet{baan08} are $\sim$0.06 and $\sim$0.04 (excluding their upper limits).

If the abundances of HCO$^{+}$ and HCN are taken to be the same throughout the Antennae and the \citet{baan08} sample, the easiest interpretation of the stronger HCO$^{+}$ emission is that the dense gas in the Antennae is less dense, on average, than in a typical FIR-luminous galaxy, where HCN (1-0) $\sim$ HCO$^{+}$ (1-0). Alternatively, differences in optical depth, which change the effective density at which the gas emits, might interact with the distribution of densities to produce stronger HCO$^{+}$ emission. Though without a functional form for the density distribution, the sense of optical depth effects is hard to quantify. Including abundance variations, there is an extensive literature on the drivers of the HCO$^{+}$-to-HCN line ratio \citep{kohno07,gracia08,krips08,baan08}. Though these differ in detail, the overall sense is that HCN emission may be enhanced relative to HCO$^+$ in the most active systems. For example, the abundance of HCO$^+$ may be additionally suppressed by recombination at high densities \citep{papadopoulos07,meier14}. In this sense, the observation of strong HCO$^{+}$ relative to HCN in the Antennae galaxies makes sense; the Antennae are extended and have low SFR-to-CO compared to more active systems \citep[see above and][]{gao01}. Indeed, the brightest HCN emission comes from the compact, active northern nucleus, where the two lines have about the same intensity. So this trend may play out within the Antennae as well as between the Antennae and other systems.

\section{Summary}

We report new CARMA observations of a suite of molecular lines in the Antennae galaxies, including the high effective density transitions HCN(1-0), HCO$^{+}$(1-0), CS(2-1), and HNC(1-0) and the bulk molecular gas tracer $^{12}$CO(1-0). We combine these with {\em Herschel} infrared observations and ALMA science verification observations of $^{12}$CO(2-1) and $^{12}$CO(3-2) to report on conditions within the molecular gas and  the apparent relationship between dense molecular gas, the total amount of molecular gas, and embedded star formation traced by infrared emission. We find substantial variations in the observed ratios IR/CO, IR/HCN, and HCN/CO across the galaxy. These variations have magnitude comparable to the scatter seen over the whole galaxy population. These measurements agree with other recent studies and highlight either substantial evolutionary effects or systematic differences in the local physical conditions in the molecular gas and its link to star formation. Comparing different molecular transitions, we find HCO$^{+}$ (1-0) to be almost twice as strong as HCN (1-0) in the overlap region, perhaps indicative of a large amount of intermediate density gas.

\section*{Acknowledgments}

We thank Christine Wilson and Fabian Walter for helpful discussions and feedback on the draft, Ulrich Klaas for assistance with the PACS data, and the referee for constructive comments that improved the draft. FB acknowledges support from DFG grant BI 1546/1-1. ER is supported by a Discovery Grant from NSERC of Canada. AU acknowledges support from Spanish MINECO grants AYA2012-32295 and FIS2012-32096. Support for CARMA construction was derived from the states of California, Illinois, and Maryland, the James S. McDonnell Foundation, the Gordon and Betty Moore Foundation, the Kenneth T. and Eileen L. Norris Foundation, the University of Chicago, the Associates of the California Institute of Technology, and the National Science Foundation. Ongoing CARMA development and operations are supported by the National Science Foundation under a cooperative agreement, and by the CARMA partner universities. This paper makes use of the following ALMA data: ADS/JAO.ALMA\#2011.0.00003.SV. ALMA is a partnership of ESO (representing its member states), NSF (USA) and NINS (Japan), together with NRC (Canada) and NSC and ASIAA (Taiwan), in cooperation with the Republic of Chile. The Joint ALMA Observatory is operated by ESO, AUI/NRAO and NAOJ. The National Radio Astronomy Observatory is a facility of the National Science Foundation operated under cooperative agreement by Associated Universities, Inc..


\begin{thebibliography}{}
\bibitem[Aniano et al.(2011)]{aniano11} Aniano, G., Draine, B.~T., Gordon, K.~D., \& Sandstrom, K.\ 2011, \pasp, 123, 1218 
\bibitem[Baan et al.(2008)]{baan08} Baan, W.~A., Henkel, C., Loenen, A.~F., Baudry, A., \& Wiklind, T.\ 2008, \aap, 477, 747 
\bibitem[Bigiel et al.(2008)]{bigiel08} Bigiel, F., Leroy, A., Walter, F., Brinks, E., de Blok, W.~J.~G., Madore, B., \& Thornley, M.~D.\ 2008, \aj, 136, 2846 
\bibitem[Bigiel et al.(2011)]{bigiel11} Bigiel, F., Leroy, A.~K., Walter, F., et al.\ 2011, \apjl, 730, L13 
\bibitem[Brandl et al.(2009)]{brandl09} Brandl, B.~R., Snijders, L., den Brok, M., et al.\ 2009, \apj, 699, 1982
\bibitem[Brogan et al.(2010)]{brogan10} Brogan, C., Johnson, K., \& Darling, J.\ 2010, \apjl, 716, L51
\bibitem[Daddi et al.(2010)]{daddi10} Daddi, E., Elbaz, D., Walter, F., et al.\ 2010, \apjl, 714, L118 
\bibitem[Evans \& Lada(1991)]{evans91} Evans, N.~J., II, \& Lada, E.~A.\ 1991, Fragmentation of Molecular Clouds and Star Formation, 147, 293 
\bibitem[Galametz et al.(2013)]{galametz13} Galametz, M., Kennicutt, R.~C., Calzetti, D., et al.\ 2013, \mnras, 431, 1956 
\bibitem[Gao et al.(2001)]{gao01} Gao, Y., Lo, K.~Y., Lee, S.-W., \& Lee, T.-H.\ 2001, \apj, 548, 172 
\bibitem[Gao \& Solomon(2004a)]{gao04a} Gao, Y., \& Solomon, P.~M.\ 2004, \apjs, 152, 63 
\bibitem[Gao \& Solomon(2004b)]{gao04b} Gao, Y., \& Solomon, P.~M.\ 2004, \apj, 606, 271 
\bibitem[Garc{\'{\i}}a-Burillo et al.(2012)]{garcia12} Garc{\'{\i}}a-Burillo, S., Usero, A., Alonso-Herrero, A., et al.\ 2012, \aap, 539, A8
\bibitem[Genzel et al.(2010)]{genzel10} Genzel, R., Tacconi, L.~J., Gracia-Carpio, J., et al.\ 2010, \mnras, 407, 2091 
\bibitem[Graci{\'a}-Carpio et al.(2008)]{gracia08} Graci{\'a}-Carpio, J., Garc{\'{\i}}a-Burillo, S., Planesas, P., Fuente, A., \& Usero, A.\ 2008, \aap, 479, 703 
\bibitem[Heiderman et al.(2010)]{heiderman10} Heiderman, A., Evans, N.~J., II, Allen, L.~E., Huard, T., \& Heyer, M.\ 2010, \apj, 723, 1019 
\bibitem[Juneau et al.(2009)]{juneau09} Juneau, S., Narayanan, D.~T., Moustakas, J., et al.\ 2009, \apj, 707, 1217
\bibitem[Kennicutt(1998)]{kennicutt98} Kennicutt, R.~C., Jr.\ 1998, \apj, 498, 541
\bibitem[Kepley(2014)]{kepley14} Kepley, A.~A., Leroy, A.~K., Frayer, D., et al.\ 2014, \apjl, 780, L13 
\bibitem[Klaas et al.(2010)]{klaas10} Klaas, U., Nielbock, M., Haas, M., Krause, O., \& Schreiber, J.\ 2010, \aap, 518, L44  
\bibitem[Kohno et al.(2007)]{kohno07} Kohno, K., Nakanishi, K., \& Imanishi, M.\ 2007, The Central Engine of Active Galactic Nuclei, 373, 647
\bibitem[Krips et al.(2008)]{krips08} Krips, M., Neri, R., Garc{\'{\i}}a-Burillo, S., et al.\ 2008, \apj, 677, 262 
\bibitem[Lada et al.(2010)]{lada10} Lada, C.~J., Lombardi, M., \& Alves, J.~F.\ 2010, \apj, 724, 687 
\bibitem[Leroy et al.(2008)]{leroy08} Leroy, A.~K., Walter, F., Brinks, E., et al.\ 2008, \aj, 136, 2782 
\bibitem[Leroy et al.(2013)]{leroy13} Leroy, A.~K., Walter, F., Sandstrom, K., et al.\ 2013, \aj, 146, 19
\bibitem[Mao et al.(2010)]{mao10} Mao, R.-Q., Schulz, A., Henkel, C., et al.\ 2010, \apj, 724, 1336 
\bibitem[Meidt et al.(2013)]{meidt13} Meidt, S.~E., Schinnerer, E., Garc{\'{\i}}a-Burillo, S., et al.\ 2013, \apj, 779, 45 
\bibitem[Meier et al.(2014)]{meier14} Meier, D.~S., Turner, J.~L., \& Beck, S.~C.\ 2014, \apj, 795, 107 
\bibitem[Murphy et al.(2011)]{murphy11} Murphy, E.~J., Condon, J.~J., Schinnerer, E., et al.\ 2011, \apj, 737, 67
\bibitem[Paglione et al.(1997)]{paglione97} Paglione, T.~A.~D., Jackson, J.~M., \& Ishizuki, S.\ 1997, \apj, 484, 656 
\bibitem[Papadopoulos(2007)]{papadopoulos07} Papadopoulos, P.~P.\ 2007, \apj, 656, 792 
\bibitem[Saintonge et al.(2011)]{saintonge11} Saintonge, A., Kauffmann, G., Wang, J., et al.\ 2011, \mnras, 415, 61  
\bibitem[Schilke et al.(1992)]{schilke92} Schilke, P., Walmsley, C.~M., Pineau Des Forets, G., et al.\ 1992, \aap, 256, 595 
\bibitem[Schirm et al.(2014)]{schirm14} Schirm, M.~R.~P., Wilson, C.~D., Parkin, T.~J., et al.\ 2014, \apj, 781, 101 
\bibitem[Schweizer et al.(2008)]{schweizer08} Schweizer, F., Burns, C.~R., Madore, B.~F., et al.\ 2008, \aj, 136, 1482
\bibitem[Sliwa et al.(2012)]{sliwa12} Sliwa, K., Wilson, C.~D., Petitpas, G.~R., et al.\ 2012, \apj, 753, 46
\bibitem[Solomon et al.(1992)]{solomon92} Solomon, P.~M., Downes, D., \& Radford, S.~J.~E.\ 1992, \apjl, 387, L55 
\bibitem[Sorai et al.(2002)]{sorai02} Sorai, K., Nakai, N., Kuno, N., \& Nishiyama, K.\ 2002, \pasj, 54, 179 
\bibitem[Stanford et al.(1990)]{stanford90} Stanford, S.~A., Sargent, A.~I., Sanders, D.~B., \& Scoville, N.~Z.\ 1990, \apj, 349, 492 
\bibitem[Ueda et al.(2012)]{ueda12} Ueda, J., Iono, D., Petitpas, G., et al.\ 2012, \apj, 745, 65 
\bibitem[van der Tak et al.(2007)]{vandertak07} van der Tak, F.~F.~S., Black, J.~H., Sch{\"o}ier, F.~L., Jansen, D.~J., \& van Dishoeck, E.~F.\ 2007, \aap, 468, 627 
\bibitem[Vigroux et al.(1996)]{vigroux96} Vigroux, L., Mirabel, F., Altieri, B., et al.\ 1996, \aap, 315, L93 
\bibitem[Wei et al.(2012)]{wei12} Wei, L.~H., Keto, E., \& Ho, L.~C.\ 2012, \apj, 750, 136 
\bibitem[Whitmore et al.(1999)]{whitmore99} Whitmore, B.~C., Zhang, Q., Leitherer, C., et al.\ 1999, \aj, 118, 1551 
\bibitem[Whitmore et al.(2010)]{whitmore10} Whitmore, B.~C., Chandar, R., Schweizer, F., et al.\ 2010, \aj, 140, 75 
\bibitem[Whitmore et al.(2014)]{whitmore14} Whitmore, B.~C., Brogan, C., Chandar, R., et al.\ 2014, \apj, 795, 156 
\bibitem[Wilson et al.(2000)]{wilson00} Wilson, C.~D., Scoville, N., Madden, S.~C., \& Charmandaris, V.\ 2000, \apj, 542, 120 
\bibitem[Wilson et al.(2003)]{wilson03} Wilson, C.~D., Scoville, N., Madden, S.~C., \& Charmandaris, V.\ 2003, \apj, 599, 1049 
\bibitem[Wilson et al.(2012)]{wilson12} Wilson, C.~D., Warren, B.~E., Israel, F.~P., et al.\ 2012, \mnras, 424, 3050 
\bibitem[Wu et al.(2005)]{wu05} Wu, J., Evans, N.~J., II, Gao, Y., et al.\ 2005, \apjl, 635, L173 
\bibitem[Wu et al.(2010)]{wu10} Wu, J., Evans, N.~J., II, Shirley, Y.~L., \& Knez, C.\ 2010, \apjs, 188, 313 
\bibitem[Yao et al.(2003)]{yao03} Yao, L., Seaquist, E.~R., Kuno, N., \& Dunne, L.\ 2003, \apj, 588, 771 
\bibitem[Zhang et al.(2010)]{zhang10} Zhang, H.-X., Gao, Y., \& Kong, X.\ 2010, \mnras, 401, 1839
\bibitem[Zhang et al.(2014)]{zhang14} Zhang, Z.-Y., Gao, Y., Henkel, C., et al.\ 2014, \apjl, 784, LL31 
\bibitem[Zhu et al.(2003)]{zhu03} Zhu, M., Seaquist, E.~R., \& Kuno, N.\ 2003, \apj, 588, 243 
\end{thebibliography}
\end{document}